\documentstyle[aps,psfig]{revtex}

\def\half{{\textstyle{1\over2}}}
\def\goes#1{\mathop{\longrightarrow}\limits_{#1}}

\def\H{{\cal H}}
\def\Z{{\cal Z}}

\def\O{{\cal O}}

\def\Vf{V_{\rm eff}}
\def\vf{v_{\rm eff}}
\def\ave#1{\left\langle#1\right\rangle}
\def\bave#1{\big\langle#1\big\rangle}
\def\dave#1{\langle\!\langle#1\rangle\!\rangle}
\def\bdave#1{\big\langle\!\!\big\langle#1\big\rangle\!\!\big\rangle}
\def\Bdave#1{\Big\langle\!\!\Big\langle#1\Big\rangle\!\!\Big\rangle}

\begin{document}

\title{Effective potential for dissipative quantum systems%
\footnote{{\em To appear on}
Proceedings of the 6th International Conference on
{\em Path-Integrals from peV TO TeV - 50 Years from Feynman's Paper},
Florence, Italy, 25-29 August 1998 (World Scientific, 1999)} }

\author{Alessandro Cuccoli, Andrea Fubini, and Valerio Tognetti}

\address{Dipartimento di Fisica dell'Universit\`a di Firenze
        and Istituto Nazionale\\ di Fisica della Materia (INFM),
        Largo E. Fermi~2, I-50125 Firenze, Italy}

\author{Andrea Rossi}

\address{Scuola Internazionale Superiore di Studi Avanzati,\\
         via Beirut 2-4, 34013 Trieste, Italy}

\author{Ruggero Vaia}

\address{Istituto di Elettronica Quantistica
         del Consiglio Nazionale delle Ricerche,\\
         via Panciatichi~56/30, I-50127 Firenze, Italy}  

\maketitle

\begin{abstract}
We consider quantum nonlinear systems with dissipation described within the
Caldeira-Leggett model, i.e., by a nonlocal action in the path integral for
the density matrix.
Approximate classical-like formulas are derived in order to evaluate thermal
averages of observables: the cases of linear and nonlinear dissipation are
considered, and the framework is extended to the case of many degrees of
freedom.
\end{abstract}
\vspace{10mm}

While several applications to condensed matter systems have demonstrated
the usefulness of the improved~\cite{GT85prl,FeynmanK86} effective potential
approach,~\cite{FeynmanH65} its suitability to treat open systems was not
immediately realized. Previous studies were basically aimed to obtain a
classical-like expression for the free energy, both in the case of strictly
linear \cite{Weiss93}, and nonlinear dissipation.~\cite{BaoZW95}

The effective-potential method, also called {\em pure-quantum self-consistent
harmonic approximation} (PQSCHA) after its generalization to phase-space
Hamiltonians~\cite{CTVV92ham,CGTVV95}, was recently used~\cite{CRTV97} to
obtain the density matrix of a nonlinear system interacting with a dissipation
bath through the Caldeira-Leggett (CL) model~\cite{CaldeiraL83a}.
In the general case the CL model starts from the Hamiltonian
\begin{equation}
 \hat\H={\hat p^2\over2m}+V(\hat q)
 +{1\over2}\sum_i\bigg\{{\hat p_i^2\over m_i} 
 + m_i\omega_i^2\Big[\hat q_i-F_i(\hat q)\Big]^2 \bigg\}~,
 \label{e.HCL}
\end{equation}
where $(\hat{p},\hat{q})$ and $(\hat{p}_i,\hat{q}_i)$ are the momenta and
coordinates of the system (here, one degree of freedom) and of an
environment (or bath) of harmonic oscillators.
The dissipation is said {\em separable} if $F_i(\hat{q})=F(\hat{q})$,
while it is said {\em linear} if also $F(\hat{q})=\hat{q}$.
In the path-integral the bath coordinates can be
integrated out exactly, leaving the CL euclidean action
\begin{equation}
 S[q(u)] =\!\int_0^{\beta} \!\!\! du
 \left[ {m\over2}\dot q^2(u) {+} V\big(q(u)\big) \right]
 -\! \int_0^{\beta} \! {du\over2} \!
 \int_0^{\beta}\!\!\!\! du'\,{\cal{R}}\big(u{-}u';q(u),q(u')\big)~.
 \label{e.S}
\end{equation}
For separable and linear dissipation we have
${\cal{R}}=K(u{-}u')\,{\cal{G}}\big(q(u),q(u')\big)$
and ${\cal{R}}=k(u{-}u')~q(u)\,q(u')$, respectively.
In the latter case, from Eq.~(\ref{e.HCL}) one can get the classical
equation of motion for the system in the Langevin form,
\[
 m\ddot q + m\int_0^\infty\!\! dt'\,\gamma(t')\,\dot q(t{-}t') + V'(q) = 0~,
\]
and the Laplace transform $\gamma(z)$ of the {\em damping function}
$\gamma(t)$ turns out to be connected with the Matsubara components
of the {\em damping kernel} $k(u)$\,:
\begin{equation}
 k(u)\equiv m\beta^{-1}~{\textstyle{\sum}_n}~e^{i\nu_nu}\,k_n~,
 ~~~~~~~~~~~ k_n=|\nu_n|\,\gamma\big(z{=}|\nu_n|\big)~,
 \label{e.kgamma}
\end{equation}
where $\nu_n=2\pi{n}/\beta$ are the Matsubara frequencies.
We will condider two cases:
\begin{eqnarray*}
 &{\rm Ohmic:~~~~~~}
 &\gamma(t)=\gamma\,\delta(t-0)~,
 ~~~~~~~\, \gamma(z)=\gamma~,
\\
 &{\rm Drude:~~~~~~}
 &\gamma(t)=\gamma\,\omega_{\rm{D}}\,e^{-\omega_{\rm{D}}t}~,
 ~~~~~~ \gamma(z)={\gamma\,\omega_{\rm{D}}/(\omega_{\rm{D}}+z)}~,
\end{eqnarray*}
where the dissipation strength $\gamma$ and the bath bandwidth
$\omega_{\rm{D}}$ characterize the environmental coupling;
$\omega_{\rm{D}}\to\infty$ gives the Ohmic (or Markovian) case.

\medskip

In the spirit of the PQSCHA, we define a trial action $S_0$ as
the most general quadratic functional with a two-times nonlocal term, i.e.,
\begin{equation}
 S_0[q(u)] = \int_0^{\beta} du
 \left[ {m\over2}\dot q^2(u) + w 
 +\half m\omega^2\,\big(q(u){-}\bar{q}\big)^2 \right]
 - \int_0^{\beta}  {du\over4}
 \int_0^{\beta} du'\,k(u{-}u')\,\big(q(u){-}q(u')\big)^2~.
 \label{e.S0}
\end{equation}
Here $\bar{q}=\beta^{-1}\int{q(u)}\,du$ is the average point of paths and
$w=w(\bar{q})$, $\omega^2=\omega^2(\bar{q})$, and $k(u)=k(u;\bar{q})$,
are parameters to be optimized by minimizing the right-hand side of
the {\em Feynman inequality}, $F\leq{F_0}-T\ave{S-S_0}_{S_0}$.

For any observable $\hat\O(\hat{p},\hat{q})$ the $S_0$-average
turns out to be expressed~\cite{CRTV97}
in terms of its Weyl symbol~\cite{Berezin80} $\O(p,q)$,
and takes the classical form
\begin{equation}
 \bave{\hat\O} ={1\over\Z_0} \sqrt{m\over{2\pi\hbar^2\beta}}
 \int\,d\bar{q}~\bdave{\O(p,\bar{q}+\xi)}~e^{-\displaystyle\beta\Vf(\bar{q})}~,
\label{e.aveOpq}
\end{equation}
where $\dave{\cdots}$ is the Gaussian average operating over $p$ and
$\xi$ with moments
\begin{eqnarray}
 \dave{\xi^2} \equiv \alpha(\bar{q})
 &=& {2\over\beta m}~ \sum\limits_{n=1}^{\infty}
 {1\over \nu_n^2+\omega^2(\bar{q})+k_n(\bar{q})}
 ~~\goes{k\to{0}}~ {1\over2m\omega}\Big(\coth f-{1\over f}\Big)
\label{e.alpha}
\\
 \dave{p^2} \equiv \lambda(\bar{q})
 &=& {m\over\beta}\,\sum\limits_{n=-\infty}^{\infty}
 {\omega^2(\bar{q})+k_n(\bar{q})
 \over \nu_n^2+\omega^2(\bar{q})+k_n(\bar{q})}
 ~~\goes{k\to{0}}~ {m\omega\over2}\coth f ~;
\label{e.lambda}
\end{eqnarray}
the effective potential is defined as
$ \Vf(\bar{q})\equiv w(\bar{q})+\sigma(\bar{q})$~, with 
\begin{equation}
 \sigma(\bar{q})={1\over\beta}\sum_{n=1}^{\infty}
 \ln\,{\nu_n^2+\omega^2(\bar{q})+k_n(\bar{q}) \over \nu_n^2}
 ~~\goes{k\to{0}}~ {1\over\beta}\ln{\sinh f\over f} ~.
\label{e.sigma}
\end{equation}
The known nondissipative limits are also reported,
with $f\equiv\beta\omega/2$.

The variational determination of the parameters gives the following
self-consistent equations:
\begin{eqnarray}
 w(\bar{q})&=&\bdave{V(\bar{q}+\xi)}-\half m\omega^2\alpha
 -{1\over2} \int_0^{\beta} du \Big[
 \Bdave{{\cal{R}}\big(u;2\bar{q}+\xi_+(u),\xi_-(u)\big)}
- {1\over2} k(u)\alpha_-(u) \Big]~,
 \label{e.w}
\\
 m\,\omega^2(\bar{q}) &=&
 \bdave{V''(\bar{q}+\xi)} - 2 \int_0^{\beta} \!\! du~
 \Bdave{\partial_{\xi_+}^2 {\cal{R}}\big(u;2\bar{q}+\xi_+(u),\xi_-(u)\big)}~,
 \label{e.o}
\\
 m\, k_n(\bar{q}) &=& \int_0^{\beta} \!\! du~
 \Bdave{\big(\partial_{\xi_+}^2{-}\partial_{\xi_-}^2\big)
 {\cal{R}}\big(u;2\bar{q}{+}\xi_+(u),\xi_-(u)\big)}~(1{-}\cos\nu_nu)\, ,
 \label{e.k}
\end{eqnarray}
where the last two arguments of ${\cal{R}}$ are to be thought as
the sum/difference variables $q_\pm(u,u')\equiv{q}(u)\pm{q}(u')$ and
\begin{equation}
 \dave{\xi^2_\pm(u)} \equiv \alpha_\pm(u)
 = {4\over\beta m}~ \sum\limits_{n=1}^{\infty}
 {1\pm\cos\nu_nu\over \nu_n^2+\omega^2(\bar{q})+k_n(\bar{q})}~.
 \label{e.alphapm}
\end{equation}

We restrict ourselves from now on to the case of linear dissipation, where
$k_n$ is given by Eq.~(\ref{e.kgamma}) and
\begin{equation}
 w(\bar{q})=\bdave{V(\bar{q}+\xi)}-\half m\omega^2\alpha~,
\hspace{10mm}
 m\,\omega^2(\bar{q}) =  \bdave{V''(\bar{q}+\xi)}~.
\end{equation}

Let us now go over to the application to a many-body system:
we consider a $\varphi^4$-chain of particles, described by the (undamped) action
\begin{equation}
 S={1\over g}\int_0^\beta du\,a\,\sum_i\left[{\dot q_i^2\over2}
 +{(q_i-q_{i-1})^2\over2a^2} + {\Omega^2\over8} \big(1-q_i^2\big)^2 \right]~,
\label{e.Sphi4}
\end{equation}
where $a$ is the chain spacing, $\Omega$ is the gap of the bare
dispersion relation, and $g$ is the quantum coupling.
The classical continuum model supports kink excitations of characteristic
width $\Omega^{-1}$ and static energy $\varepsilon_{\rm{K}}=2\Omega/3g$,
which is naturally used as the energy scale in defining the reduced
temperature $t\equiv(\beta\varepsilon_{\rm{K}})^{-1}$; an alternative
coupling constant is defined as the ratio between the quantum harmonic
energy $\Omega$ and $\varepsilon_{\rm{K}}$: $Q\equiv{3g/2}$.
We also use the kink length in lattice units $R\equiv(a\Omega)^{-1}$
($R\to\infty$ in the continuum limit).
We assume uncorrelated identical CL baths for each degree of freedom,
and use the low-coupling approximation~\cite{CGTVV95} in order
to deal with the effective potential, by which the partition function
can be eventually written as
\begin{eqnarray}
 {\cal{Z}}&=& \bigg({3t\over4\pi RQ^2}\bigg)^{N\over2}\!\!\!\int\! d^N \! q
 \exp\bigg\{\!-{1\over Rt} {\sum}_i\Big[{R^2\over2} (q_i{-}q_{i-1})^2
 + \vf(q_i) \Big]\!\bigg\} \, ,
\label{e.Vfphi4}
\\
 & &~~~~~ \vf(q)={1\over8}\big(1-3D-q^2\big)^2 + {3\over4}D^2
 +{Rt\over N}{\sum}_k\sigma_k~.
\label{e.vfphi4}
\end{eqnarray}
The renormalization parameter $D=D(t;Q,R;\gamma,\omega_{\rm{D}})$
generalizes Eq.~(\ref{e.alpha}) and is the solution of the self-consistent
equations
\begin{eqnarray}
 D&=&{4Rt\over3N}{\sum}_k{\sum}_n {\Omega^2\over\nu_n^2+\omega_k^2+k_n},
\\
 \omega_k^2(t)&=&\Omega^2\big[(1-3D)+4R^2\sin^2(ka/2)\big]~,
\end{eqnarray}
where $k_n=\gamma\omega_{\rm{D}}\,\nu_n/(\omega_{\rm{D}}+\nu_n)$ and
here $\nu_n/\Omega=2\pi{n}t/Q$.

We see that while the averages of coordinate-dependent observables tend
to the classical behavior (in other words, the environment quenches the
pure-quantum coordinate fluctuations), those of momentum-dependent ones are
enhanced due to the momentum exchanges with the environment,
resulting in unexpected behaviour for ``mixed'' quantities like the
specific heat.

\newpage

\begin{figure}[h]
\centerline{
\begin{minipage}[t]{85mm}
\centerline{\psfig{bbllx=9mm,bblly=60mm,bburx=194mm,bbury=256mm,%
figure=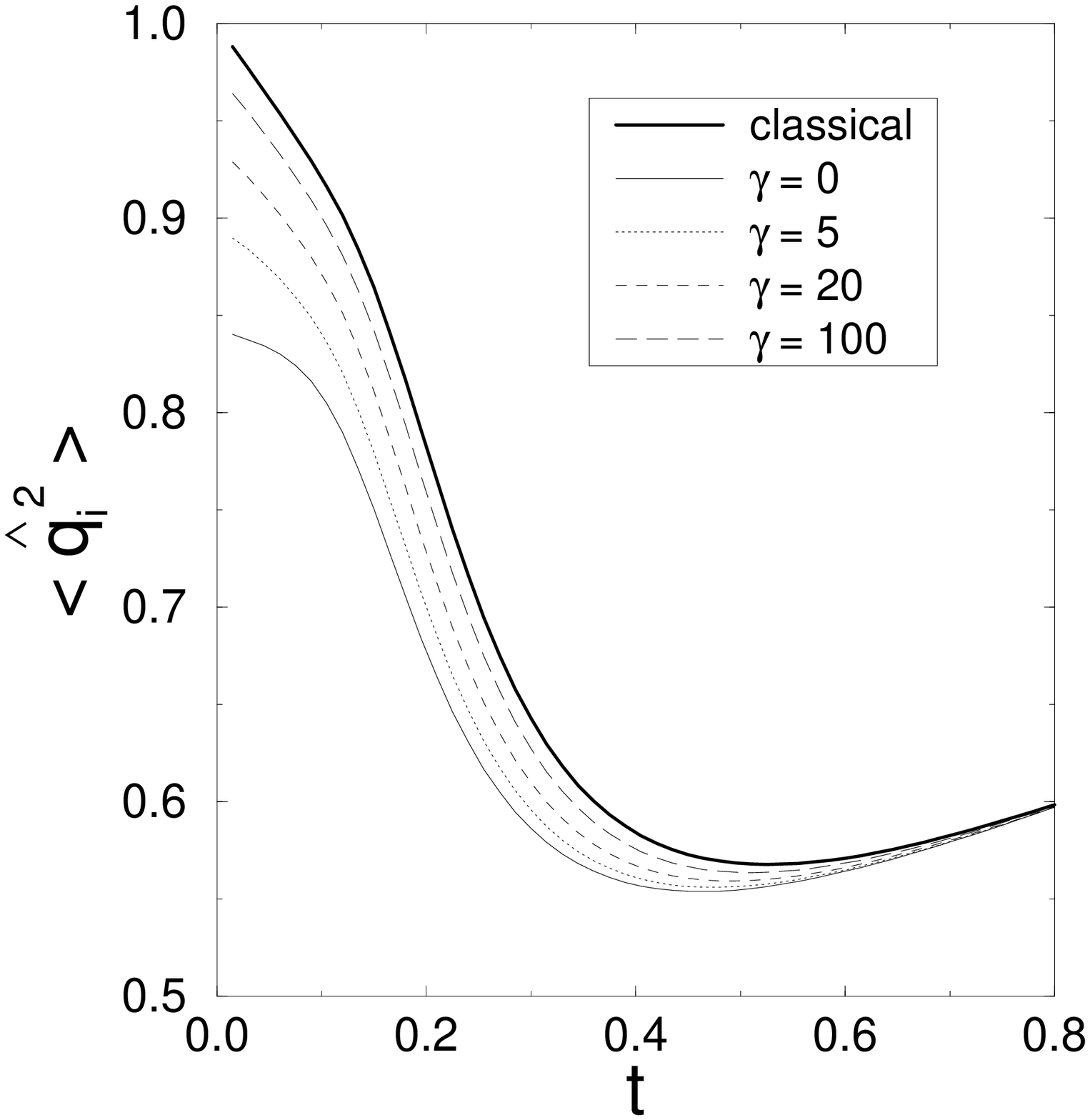,width=85mm,angle=0}}
\end{minipage}
\hfill
\begin{minipage}[t]{85mm}
\centerline{\psfig{bbllx=9mm,bblly=60mm,bburx=194mm,bbury=256mm,%
figure=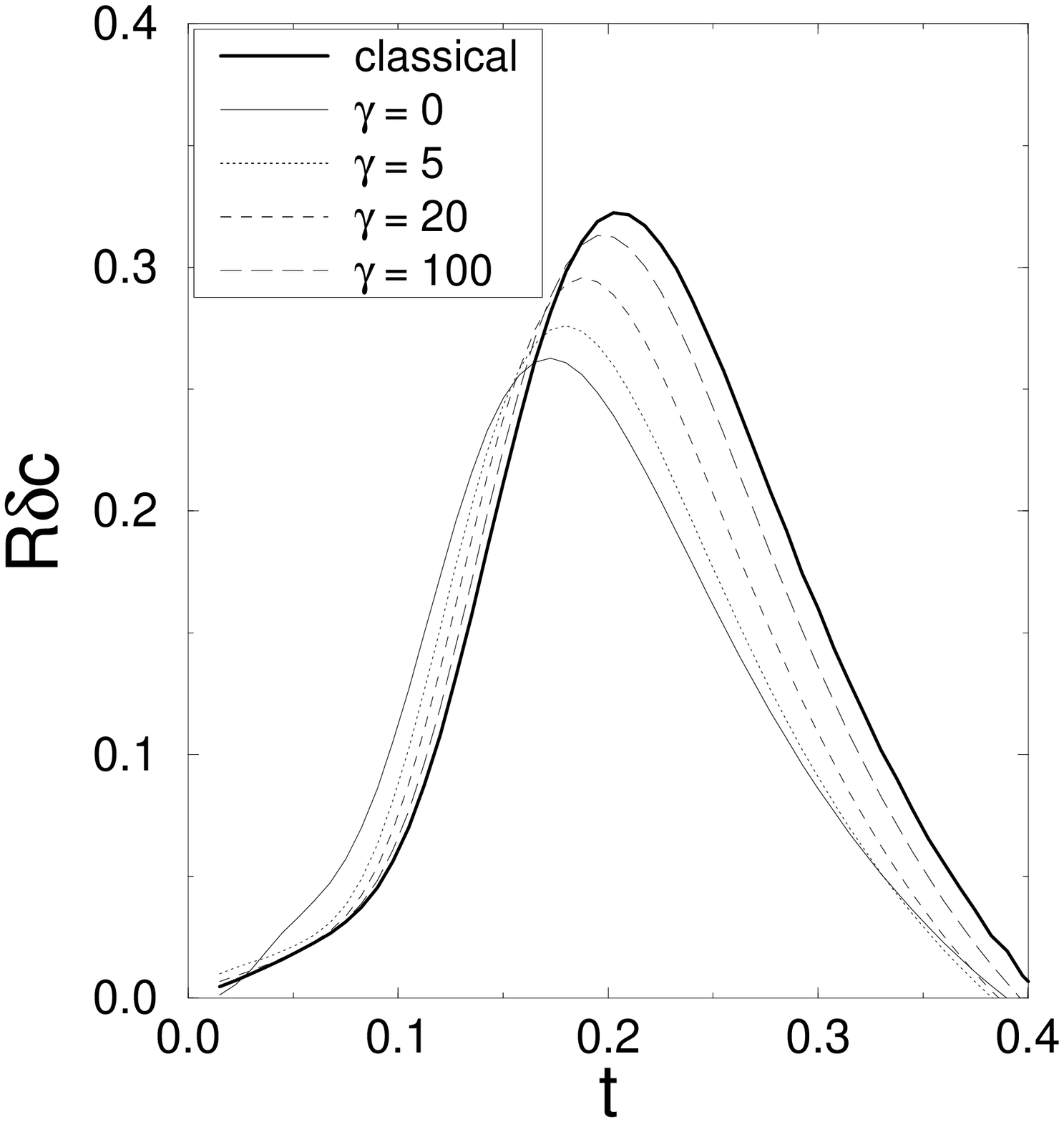,width=85mm,angle=0}}
\end{minipage}
}
\caption{Double-well $\varphi^4$-chain:
mean square fluctuation of the site coordinate $\langle\hat{q}_i^2\rangle$
vs. $t$ (left),
and nonlinear specific heat $\delta c(t)=c(t){-}c_{\rm{h}}(t)$ (right), 
for $Q=0.2$, $R=5$. The curves refer to different Drude dissipation
strengths $\gamma$ (in units of $\Omega$), with
$\omega_{\rm{D}}/\Omega=100$. Both quantities tend to the classical
behaviour for increasing dissipation strength: this is expected for
$\langle\hat{q}_i^2\rangle$, while it is a non trivial
result for $\delta c(t)=c(t){-}c_{\rm{h}}(t)$.
\label{f.1}
} 
\end{figure}

\vspace{10mm}

\begin{figure}[h]
\centerline{
\begin{minipage}[t]{85mm}
\centerline{\psfig{bbllx=14mm,bblly=60mm,bburx=194mm,bbury=235mm,%
figure=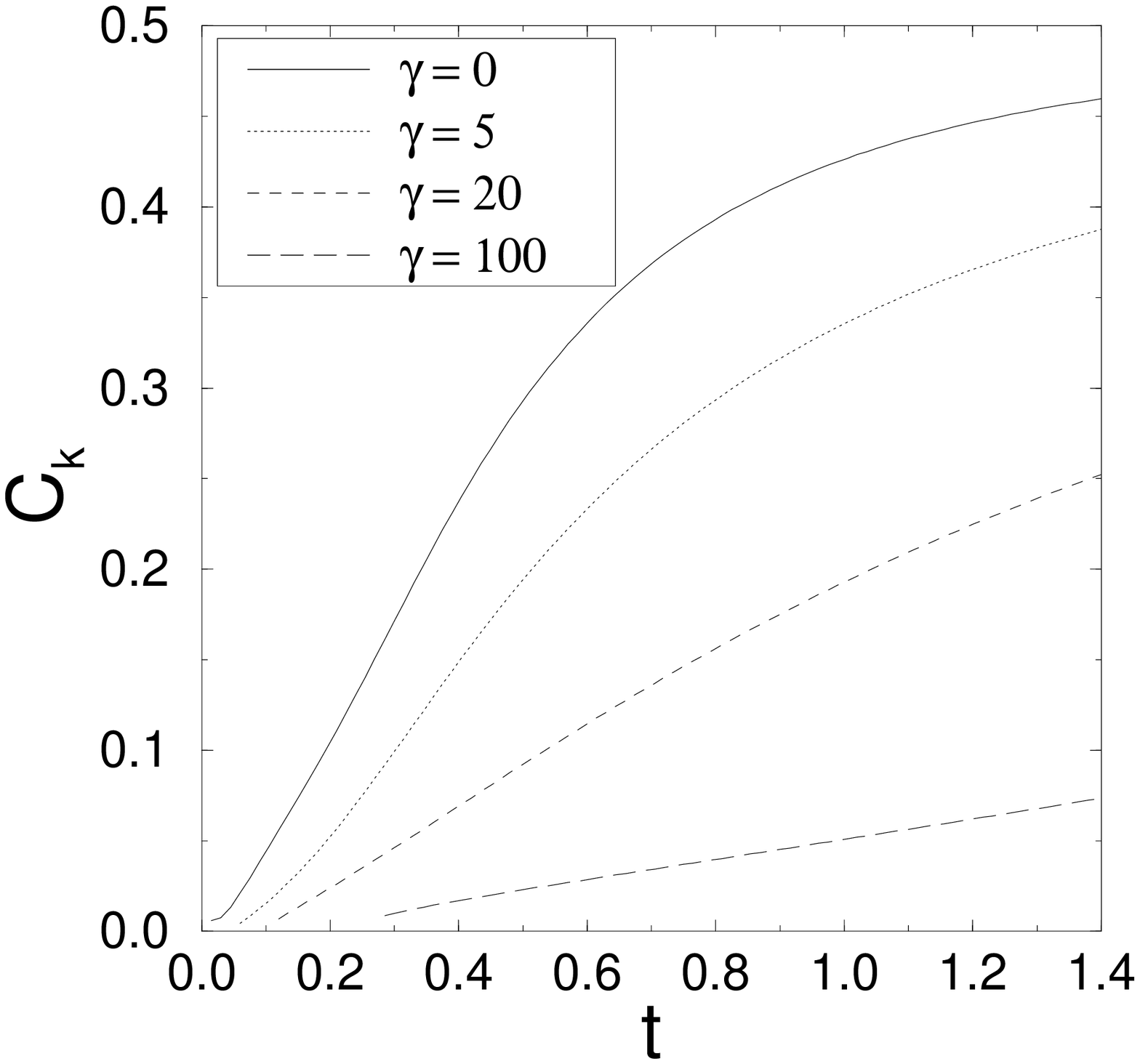,width=85mm,angle=0}}
\end{minipage}
\hfill
\begin{minipage}[t]{85mm}
\centerline{\psfig{bbllx=14mm,bblly=60mm,bburx=194mm,bbury=235mm,%
figure=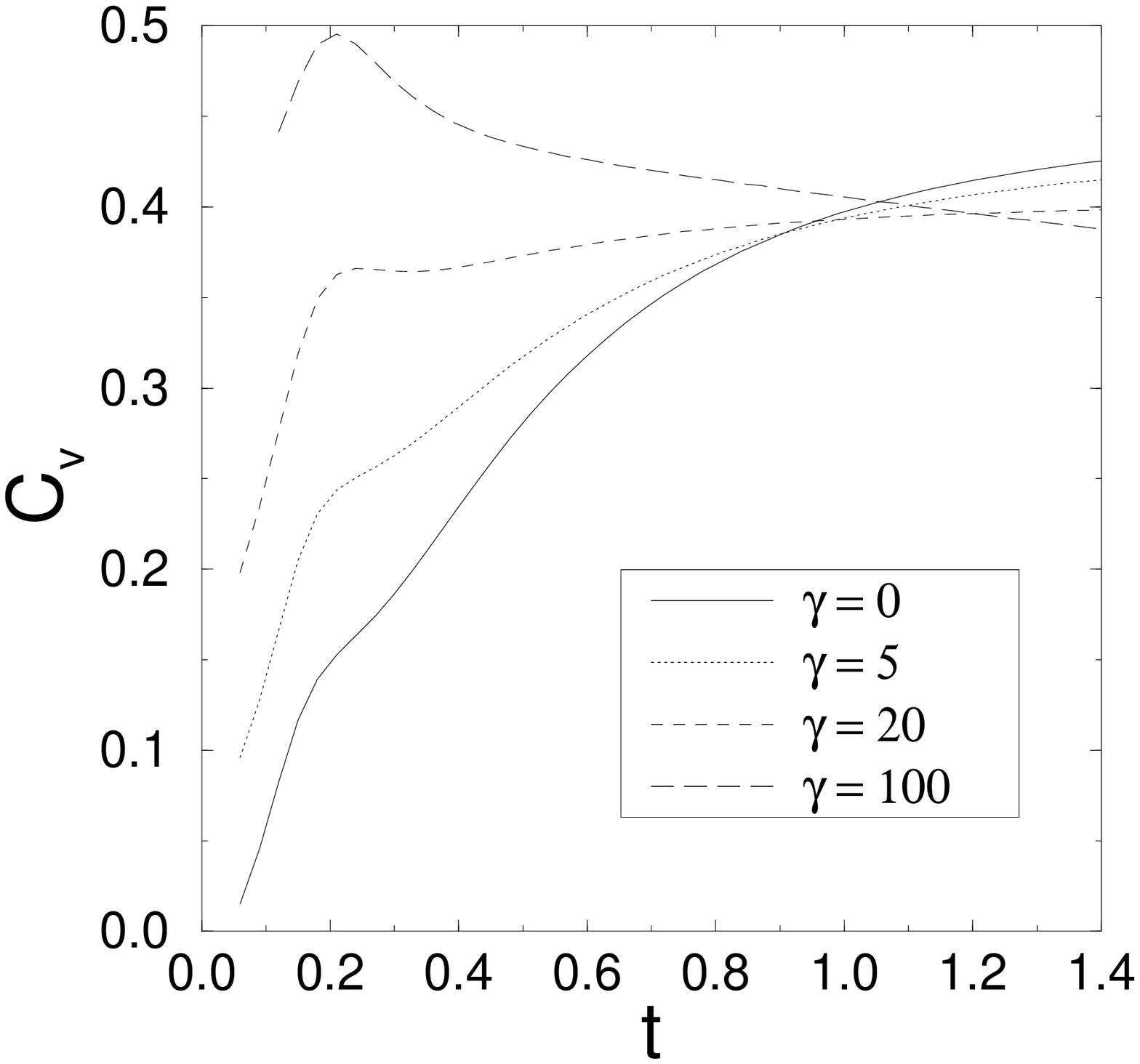,width=85mm,angle=0}}
\end{minipage}
}
\caption{Double-well $\varphi^4$-chain: kinetic (left) and interaction
(right) parts of the specific heat, $c_{\scriptscriptstyle{\rm{K}}}(t)$
and $c_{\scriptscriptstyle{\rm{V}}}(t)$, for $Q=0.2$, $R=5$.
The curves refer to different Drude dissipation
strengths $\gamma$ (in units of $\Omega$), with
$\omega_{\rm{D}}/\Omega=100$.
The kinetic part appears to be quenched by dissipation: the kinetic
energy (substantially $\langle\hat{p}_i^2\rangle$) becomes indeed larger,
but is flatter in $t$, so its derivative $c_{\scriptscriptstyle{\rm{K}}}(t)$
decreases with $\gamma$.
\label{f.2}
} 
\end{figure}

\end{document}